\documentclass[showpacs,superscriptaddress,preprintnumbers,amsmath,amssymb,twocolumn,nofootinbib]{revtex4}

\usepackage{graphicx}
\usepackage{dcolumn}
\usepackage{bm}
\usepackage{type1cm}

\newcommand{\phizero}{\!{\stackrel{\,_{(0)}}\phi}\!}

\newcommand{\varphizero}{\!{\stackrel{\,_{(0)}}\varphi}\!}
\newcommand{\varphione}{\!{\stackrel{\,_{(1)}}\varphi}\!}
\newcommand{\varphitwo}{\!{\stackrel{\,_{(2)}}\varphi}\!}
\newcommand{\tvarphione}{\!{\stackrel{\,_{(1)}}{\tilde \varphi}}\!}
\newcommand{\tvarphitwo}{\!{\stackrel{\,_{(2)}}{\tilde \varphi}}\!}

\begin{document}

\preprint{KUNS-2110}

\title{Primordial Non-Gaussianity in Multi-Scalar Inflation}

\author{Shuichiro Yokoyama}
\email{shu@tap.scphys.kyoto-u.ac.jp}
\affiliation{Department of Physics, Kyoto University, Kyoto 606-8502, Japan}

\author{Teruaki Suyama}%
\email{suyama@icrr.u-tokyo.ac.jp}
\affiliation{Institute for Cosmic Ray Research, The University of Tokyo, Kashiwa 277-8582, Japan}

\author{Takahiro Tanaka}
\email{tama@scphys.kyoto-u.ac.jp}
\affiliation{Department of Physics, Kyoto University, Kyoto 606-8502, Japan}
 
\date{\today}

\begin{abstract}
We give a concise formula for the non-Gaussianity of the primordial 
curvature perturbation generated on super-horizon scales in multi-scalar 
inflation model without assuming slow-roll conditions. 
This is an extension of our previous work.
Using this formula, 
we study the generation of non-Gaussianity for the double inflation 
models in which the slow-roll conditions are temporarily violated 
after horizon exit, 
and we show that the non-linear parameter $f_{NL}$ for such models is 
suppressed by the slow-roll parameters evaluated at the time of horizon exit.


\end{abstract}

\pacs{98.80.Bp, 98.80.Cq, 98.80.Es}
\maketitle
\section{Introduction}
Non-Gaussianity of the primordial curvature perturbation is a potentially useful 
discriminator of the many existing inflation models \cite{Komatsu:2001rj,Bartolo:2004if}.
Planck \cite{:2006uk} is expected to detect the primordial non-Gaussianity
if the so-called non-linear parameter, $f_{NL}$, which parameterizes the 
magnitude of the bispectrum, is larger than $3\sim 5$ \cite{Komatsu:2001rj, Babich:2004yc}.
Higher order correlation functions such as trispectrum would also be 
a useful probe of the primordial non-Gaussianity \cite{Okamoto:2002ik, Bartolo:2005fp, Kogo:2006kh}.
Hence it is important to theoretically understand the generation of non-Gaussianity.

Standard single slow-roll inflation model predicts rather small level of
the non-linear parameter, $f_{NL}$,
suppressed by the  slow-roll parameters \cite{Maldacena:2002vr}.
In multi-scalar field inflation models with separable potential,
explicit calculations in the slow-roll approximation show that $f_{NL}$ is also of the order of 
the slow-roll parameters~\cite{Kim:2006te,Vernizzi:2006ve,Battefeld:2006sz,Battefeld:2007en,Choi:2007su,Rigopoulos:2005xx,Rigopoulos:2005us,Rigopoulos:2005ae}.
This feature is likely to hold for more general potential that
satisfies the slow-roll conditions \cite{Seery:2005wm,Seery:2005gb,Yokoyama:2007uu}.
On the other hand, 
the generation of non-Gaussianity due to the violation 
of the slow-roll conditions remains an open problem.

In this paper, 
based on $\delta N$ formalism~\cite{Sasaki:1995aw,Sasaki:1998ug,Lyth:2005fi,Lyth:2004gb,Starobinsky:1986fx},
we give a useful formula for calculating the non-linear parameter 
in the multi-scalar inflation models including the models in which the 
slow-roll approximation is (temporarily) violated after the cosmological 
scales exit the horizon scale during inflation. 
Current observations do not exclude such models.
We also study the possibility of the large non-Gaussianity for double 
inflation model~\cite{Silk:1986vc,Turner:1986te}, 
as an example of such a model. 

In section \ref{formulation}, 
we derive the formula for the non-linear parameter $f_{NL}$,
which is an extension of our previous paper~\cite{Yokoyama:2007uu}. 
In section \ref{example}, 
we analyze the primordial non-Gaussianity in the double inflation model
using our formula. 
We give a summary in section \ref{summary}.

\section{formulation}
\label{formulation}
 In this section, we derive a formula which is an extension of that given in our previous paper \cite{Yokoyama:2007uu} to the non slow-roll case. We use the unit $M_{\rm pl}^2=\left(8\pi G\right)^{-1}=1$.
\subsection{Background equations}

We consider a ${\cal N}$-component scalar field whose action is given by
\begin{eqnarray}
S = - \int d^4 x \sqrt{-g}\left[{1 \over 2}h_{IJ}g^{\mu\nu}\partial_{\mu}\phi^I\partial_{\nu}\phi^J + V(\phi)\right]~,&&\\
\left(I,J=1,2,\cdots,{\cal N}\right)~,&&\nonumber
\end{eqnarray}
where $g_{\mu\nu}$ is the spacetime metric and $h_{IJ}$ is the metric on the scalar field space. 
In the main text,
we restrict our discussion to the flat field space metric $h_{IJ}=\delta_{IJ}$
to avoid inessential complexities due to non-flat field space metric.
Extension to the general field space metric is given in appendix \ref{hij}.

We define $\varphi^I_{i}(i=1,2)$ as
\begin{eqnarray}
\varphi^I_1 \equiv \phi^I~,~~\varphi^I_2 \equiv \frac{d}{dN}\phi^I~,
\end{eqnarray}
where $dN=Hdt$ with $H$ and $t$ being the Hubble parameter and cosmological time, respectively. 
Namely, we take $e$-folding number, $N$, as a time coordinate.
For brevity, hereinafter, we use Latin indices at the beginning of Latin alphabet, $a$, $b$ or $c$, instead of the double indices, $I,~i$, i.e., $X^a =X^I_i$.

Then, 
the background equation of motion for $\varphi^a$ is
\begin{eqnarray}
{d \over dN}\varphi^a = F^a(\varphi)~, \label{back1}
\end{eqnarray}
where $F^a (= F^I_i)$ is given by  
\begin{eqnarray}
F^I_1 = \varphi^I_2~,~~F^I_2 = - {V \over H^2}\left( \varphi^I_2 + {V^I \over V}\right)~,
\end{eqnarray}
where $V^I=\delta^{IJ}(\partial V/\partial\phi^I)$, 
and the Friedmann equation is
\begin{equation}
H^2 = {2V \over 6 - \varphi^I_2 \varphi_{2I}}~, \label{friedmann}
\end{equation}
with $\varphi_{2I} = \delta_{IJ}\varphi^J_2$.

\subsection{Perturbations}
\label{perturbations}
In the $\delta N$ formalism~\cite{Sasaki:1995aw,Sasaki:1998ug,Lyth:2005fi},
 the evolution of the difference between two
adjacent background solutions determines that of the
primordial curvature perturbation on super-horizon scales.
In this paper,
we use the word "perturbation" to denote the difference between 
two adjacent background solutions.
In this subsection, we analyze the time evolution of the perturbation
and relate the result to the curvature perturbation.

The solution of the background equation~(\ref{back1})
is labelled by $2{\cal N}$ integral constants $\lambda^a$.
Let us define $\delta \varphi^a$ as the perturbation,
\begin{eqnarray}
\delta \varphi^a (N)  \equiv \varphi^a( \lambda+\delta\lambda; N)-\varphi^a(\lambda; N)~, \label{per1}
\end{eqnarray}
where $\lambda$ is abbreviation of $\lambda^a$
and $\delta\lambda^a$ is a small quantity of ${\cal O}(\delta)$.
$\delta \varphi^a (N)$ defined by Eq.~(\ref{per1}) represents a perturbation of the scalar field on the
$N= {\rm constant}$ gauge~\cite{Sasaki:1998ug}.
For the purpose of calculating the leading bispectrum
of the curvature perturbation,
it is enough to know the evolution of $\delta \varphi^a (N)$
up to second order in $\delta$.
For later convenience,
we decompose $\delta \varphi^a$ as
\begin{eqnarray}
\delta \varphi^a=\delta \varphione^a+{1 \over 2}\delta \varphitwo^a~,
\end{eqnarray}
where $\delta \varphione^a$ and $\delta \varphitwo^a$
are first and second order quantities in $\delta$,
respectively.

Evolution equation for $\delta \varphione^a$ is given by
\begin{eqnarray}
{d \over dN}\delta \varphione^a (N) = P^{a}_{~b}(N)  \delta \varphione^b(N)~, 
\label{first}
\end{eqnarray}
where ${P}^{a}_{~b} (= P^{Ij}_{iJ})$ is defined by
\begin{eqnarray}
\left.P^{a}_{~b} \equiv {\partial F^a \over \partial
 \varphi^b}\right\vert_{\varphi=\varphizero(N)}~.
\label{Pdefined}
\end{eqnarray}
Here $\varphizero(N)$ represents the unperturbed trajectory. 
The explicit form of $P^{a}_{~b}$ is shown in appendix \ref{specific}.
Formally,
solution of this equation can be written as
\begin{eqnarray}
\delta \varphione^a(N) = \Lambda^{a}_{~b}(N,N_*)\delta \varphione^b(N_*)~, \label{sol1}
\end{eqnarray}
where $\Lambda^{a}_{~b}$ is a solution of 
\begin{eqnarray}
{d \over dN}\Lambda^{a}_{~b}(N,N') = P^{a}_{~c}(N)\Lambda^{c}_{~b}(N,N')~,
\end{eqnarray}
with the condition $\Lambda^{a}_{~b}(N,N)=\Lambda^{Ij}_{iJ}(N,N) =
\delta^I_{~J} \delta^{~j}_i$. 

Evolution equation for $\delta \varphitwo^a$ is given by
\begin{eqnarray}
{d \over dN}\delta \varphitwo^a(N) & =  & {P}^{a}_{~b}(N)\delta
 \varphitwo^b(N) 
\cr &&\quad + Q^{a}_{~bc}(N)\delta 
\varphione^b(N)\delta \varphione^c(N)~,
\label{second}
\end{eqnarray}
where $Q^{a}_{~bc}$ is defined by
\begin{eqnarray}
Q^{a}_{~bc} \equiv {\partial^2 F^a \over \partial\varphi^b \partial\varphi^c}\biggr|_{\varphi=\varphizero (N)} 
= {\partial P^{a}_{~b}\over \partial \varphi^c}\biggr|_{\varphi=\varphizero (N)}~.
\label{Qdefined}
\end{eqnarray}
The explicit form of $Q^{a}_{~bc}( = Q^{Ijk}_{iJK})$ is shown in appendix \ref{specific}.
Let us choose the integral constants $\lambda^a$ as the 
initial values of $\varphi^a$ at $N=N_*$,
namely,
$\lambda^a=\varphi^a(N_*)$.
Then we have $\delta \varphi^a(N_*)=\delta \lambda^a$.
Hence $\delta \varphitwo^a(N)$ vanishes at $N_*$. 
Under this initial condition,
the formal solution of Eq.~(\ref{second}) is given by
\begin{eqnarray}
\delta \varphitwo^a(N) &=&\int^N_{N_*}dN'\Lambda^{a}_{~b}(N,N')
Q^{b}_{~cd}(N') \nonumber\\
&&\qquad\qquad\quad \times\delta \varphione^c(N')\delta \varphione^d(N')~.\label{sol2}
\end{eqnarray}

According to the $\delta N$ formalism~\cite{Sasaki:1995aw,Lyth:2005fi}, 
the curvature perturbation on large scales evaluated at a final time, $N=N_c$,
is given by the perturbation of the $e$-folding number 
between an initial flat hypersurface at $N=N_*$ and a final uniform energy density hypersurface 
at $N=N_c$.  
Let us take $N_*$ to be a certain time soon after the relevant length scale crossed the horizon scale, $H^{-1}$, during the scalar 
dominant phase and 
$N_c$ to be a certain time after the complete convergence of the background trajectories has occurred.
At $N > N_c$ the dynamics of the universe is characterized by a single parameter and only the adiabatic perturbations remain\footnote{
We consider the case in which isocurvature perturbations do not persist until later.
}.
Then, the $e$-folding number between $N_*$ and $N_c$ can be regarded as the function of the final time $N_c$ 
and $\varphi^a(N_*)$, which we denote $N(N_c, \varphi(N_*))$. 
  
Based on $\delta N$ formalism, the curvature perturbation on the uniform energy density hypersurface evaluated at $N=N_c$ is given by
\begin{eqnarray}
\zeta(N_c) &\simeq& \delta N (N_c, \varphi(N_*)) \nonumber\\
&=& N_{a*}\delta \varphi^{a}_{*} + {1 \over 2}N_{ab*}\delta \varphi^a_{*} \delta \varphi^b_{*} + 
\cdots
~, \label{nl1}
\end{eqnarray}
where $\delta \varphi^a_{*} = \delta \varphi^a(N_*)$ represents the
field perturbations and their time derivative on the initial flat
hypersurface at $N=N_*$. 
The left hand side in Eq.~(\ref{nl1}) is obviously independent of the initial time $N_*$,
and hence so is $\delta N(N_c, \varphi(N_*))$.
Here we also defined $N_{a*} = N_a(N_*)$ and $N_{ab*} = N_{ab}(N_*)$ 
by 
\begin{eqnarray}
N_{a}(N) &\equiv& {\partial N (N_c,\varphi) \over \partial\varphi^a}
\biggr|_{\varphi=\varphizero (N)}~,\\
N_{ab}(N) &\equiv& {\partial^2 N (N_c,\varphi)\over \partial \varphi^a \partial\varphi^b}
\biggr|_{\varphi=\varphizero (N)}~,  
\end{eqnarray}
evaluated at $N=N_*$.  

It is well known that the curvature perturbations on an uniform density hypersurface, $\zeta$, remain constant in time 
for $N > N_c$.
Hence, $\zeta(N_c)$ gives the final spectrum of the primordial perturbation.

Let us take $N_F$ to be a certain late time during the scalar dominant phase. Then we have
\begin{eqnarray}
\zeta(N_c) &\simeq& \delta N (N_c, \varphi(N_F)) \nonumber\\
 &=& N_{aF}\delta \varphi^a_{F} + {1 \over 2}N_{abF}\delta \varphi^a_{F} \delta \varphi^b_{F} + 
 \cdots
 ~. \label{nl2}
\end{eqnarray}
where $\delta \varphi^a_{F} = \delta \varphi^a(N_F)$, $N_{aF}=N_a(N_F)$ and $N_{abF} = N_{ab}(N_F)$.
During the period with $N_*<N<N_F$, 
we can use the solutions for $\delta \varphi^a$ 
given by Eqs.~(\ref{sol1}) and (\ref{sol2}).
Using these solutions, 
we obtain the relations;
\begin{eqnarray}
&&N_{a*} = N_{bF}\Lambda^{b}_{~a}(N_F,N_*)~,\\
&&N_{ab*} = N_{cdF}\Lambda^{c}_{~a}(N_F,N_*)\Lambda^{d}_{~b}(N_F,N_*) \nonumber\\
&& \qquad\qquad
+2\int^{N_F}_{N_*}dN' N_c(N')Q^{c}_{~de}(N') \nonumber\\
&& \qquad\qquad\qquad\quad
\times \Lambda^{d}_{~a}(N',N_*)\Lambda^{e}_{~b}(N',N_*)~,
\end{eqnarray}
with 
\begin{eqnarray}
N_a (N) \equiv N_{bF} \Lambda^{b}_{~a} (N_F,N).\label{defN}
\end{eqnarray}

\subsection{Non-linear parameter}
In this subsection,
we derive a formula for the non-linear parameter $f_{NL}$ by
making use of the $\delta N$ formalism.
We first give the definition of $f_{NL}$.
It is defined as the magnitude of the bispectrum
of the curvature perturbation $\zeta$,
\begin{eqnarray}
\!\! B_\zeta(k_1,k_2,k_3)&\!\! =&\!\! \frac{6}{5} \frac{f_{NL}}{{(2\pi)}^{3/2}} 
\biggl[ P_\zeta (k_1) P_\zeta(k_2)
   \cr &&
   + P_\zeta (k_2) P_\zeta(k_3)
   +P_\zeta (k_3) P_\zeta(k_1) \biggr]~,\nonumber\\ 
\label{defi}
\end{eqnarray}
where $P_\zeta$ is the power spectrum of $\zeta$.
The definitions of $P_\zeta$ and $B_{\zeta}$ are, respectively, 
\begin{eqnarray}
&&\langle \zeta_{{\bf k}_1} \zeta_{{\bf k}_2} \rangle \equiv \delta ({\bf k}_1+{\bf k}_2 ) P_\zeta (k_1)~, \\
&&\langle \zeta_{{\bf k}_1} \zeta_{{\bf k}_2} \zeta_{{\bf k}_3} \rangle \equiv \delta ({\bf k}_1+{\bf k}_2+{\bf k}_3 )B_\zeta (k_1,k_2,k_3)~.
\end{eqnarray}
Equation~(\ref{defi}) restricts the form of the bispectrum.
The bispectrum in general does not take that simple form.
In fact, sub-horizon perturbations of fields give different
$k$-dependent form of the bispectrum \cite{Seery:2005wm}.
However,
the sub-horizon contribution to the bispectrum is suppressed
by the slow-roll parameters evaluated at the time of horizon exit\footnote{
Here, we consider the case that inflation is induced by the canonical scalar fields.
In such a case, current observations of the spectrum of curvature perturbation restrict 
 the models so that the slow-roll conditions are satisfied until the cosmologically
relevant scales exit of the horizon scale. 
Otherwise, for example, in Dirac-Born-Infeld (DBI) inflation model, the sub-horizon contribution to the bispectrum not only has different $k$-dependent form,
but also can be large enough to be detectable in the future experiment without inconsistency with current observations
~\cite{Chen:2006nt}. 
}. 
In contrast, 
the super-horizon evolution always gives the bispectrum in the form of
Eq.~(\ref{defi}) independent of the number of fields (see below).
If $f_{NL} \gtrsim 1$,
which is an interesting case from the observational point of view,
then the contribution due to super-horizon evolution dominates the total bispectrum.

We assume that the slow-roll conditions are satisfied at $N=N_*$.
Then, 
to a good approximation,
$\delta \varphi^I_{1*}$ becomes a Gaussian variable~\cite{Maldacena:2002vr,Seery:2005wm} 
with its variance given by 
\begin{eqnarray}
\langle \delta \varphi^I_{1*} \delta \varphi^J_{1*} \rangle= \delta^{IJ}{\left( \frac{H_\ast}{2\pi} \right)}^2~,
\end{eqnarray}
 and
$\varphi^I_2$ becomes function of $\varphi^I_1$.
Differentiating $\varphi^I_2 \simeq -\frac{V^I}{V}$,
we have
\begin{eqnarray}
\delta \varphi^I_{2*} =\left( \frac{V^I V_J}{V^2}-\frac{V^{I}_{~J}}{V} \right) \delta \varphi^J_{1*}+\cdots. 
\end{eqnarray}
The higher order terms are also suppressed by the slow-roll
parameters.
Hence, 
$\delta \varphi^I_{2*}$ is Gaussian as is $\delta \varphi^I_{1*}$ 
to a good approximation.
Then, 
we can write down the variance of $\delta \varphi^a_*$ as 
\begin{eqnarray}
\langle \delta \varphi^a_* \delta \varphi^b_*\rangle \simeq A^{ab} 
{\left( \frac{H_\ast}{2\pi} \right)}^2~.
\end{eqnarray}
At the first order both in the field perturbation and slow-roll limit,
the matrix $A^{ab} = A^{IJ}_{ij}$ can be written as
\begin{eqnarray}
A^{IJ}_{11} &\!\! =&\!\!\delta^{IJ}~,~~~A^{IJ}_{12}=A^{IJ}_{21}=\epsilon^{IJ}~, \nonumber\\
A^{IJ}_{22}&\!\! =&\!\! \epsilon^{I}_{\,K} \epsilon^{KJ}~,
 \label{aij}
\end{eqnarray}
where
\begin{eqnarray}
\epsilon^{IJ} \equiv \left[\frac{V^I(\phi) V^J(\phi)}{V(\phi)^2}-\frac{V^{IJ}(\phi)}{V(\phi)}~\right]_{\phi=\phizero (N_*)}~. 
\end{eqnarray}
 Since $\epsilon^{IJ} = O(\epsilon,\eta)$, we find that $\langle \delta \varphi^I_{1*} \delta \varphi^J_{2*} \rangle$ and 
$\langle \delta \varphi^I_{2*} \delta \varphi^J_{2*} \rangle$ are suppressed by the slow-roll parameters.
At the same order, $O(\epsilon,\eta)$, it is known that we need to add the slow-roll correction terms to $A^{IJ}_{11}$ in Eq.~(\ref{aij}). 
Such corrections to $A^{IJ}_{11}$ are given in \cite{Byrnes:2006vq}.

Using these equations, to the leading order,
the non-linear parameter is written as
\begin{eqnarray}
{6 \over 5}f_{NL} &\!\! \simeq& \!\! {N_{a*}N_{b*}N_{cd*}A^{ac} A^{bd}
\over \left(N_{e*}N_{f*}A^{ef}\right)^2} \nonumber \\
&=&\!\! {1\over \left(N_{a*}\Theta^{a}_{*}\right)^{2}}
\Bigg[  N_{abF}\Theta^a(N_F)\Theta^b(N_F) \nonumber \\
 &&\hspace{8mm}+  \int^{N_F}_{N_*}\!\! dN'N_c(N')Q^{c}_{~ab}(N') \nonumber\\
 &&\hspace{25mm} \times \Theta^{a}(N')\Theta^{b}(N') \Bigg]~, 
\label{fnl}
\end{eqnarray}
where 
\begin{eqnarray}
\Theta^a(N) \equiv \Lambda^{a}_{~c}(N,N_*)A^{cb}N_{b*}~,
\label{deftheta}
\end{eqnarray}
and $\Theta^a_* = \Theta^a(N_*)$.
As we mentioned before, we have neglected the non-Gaussianity from the sub-horizon contributions in deriving Eq.~(\ref{fnl}).
Eq.~(\ref{fnl}) shows that,
aside from $N_{aF}$ and $N_{abF}$ whose explicit
form will be given in the next subsection \ref{expression},
$f_{NL}$ is completely determined by the quantities 
$N_a (N)$ and $\Theta^a (N)$.
These quantities obey the following closed differential equations,
\begin{eqnarray}
&&{d \over dN}N_a(N) = - N_b(N)P^{b}_{~a}(N)~, \label{diff1} \\
&&{d \over dN}\Theta^{a}(N) = P^{a}_{~b}(N) \Theta^{b}(N)~ \label{diff2}.
\end{eqnarray}
First,
we solve Eq.~(\ref{diff1}) backward till $N=N_\ast$
under the initial conditions $N_a(N_F)=N_{aF}$.
Then we solve Eq.~(\ref{diff2}) forward till $N=N_F$
under the initial conditions 
$\Theta^a (N_\ast)= A^{ab}N_{b\ast} $. 
Substituting these solutions into Eq.~(\ref{fnl}),
we obtain $f_{NL}$.


The evolution equation of $\Theta^a(N)$, Eq.~(\ref{diff2}), is identical to
that of $\delta \varphi^a$
, Eq.~(\ref{first}), at the linear level.
Since Eqs.~(\ref{diff1}) and (\ref{diff2}) are mutually dual,
a variable composed of $N_a(N_*)$ and $\Theta^a(N_*)$;
\begin{eqnarray}
W(N_*) = N_a(N_*) \Theta^a(N_*)~,\label{wrons}
\end{eqnarray}
becomes constant irrespective of $N_*$.
This constancy of $W(N_*)$ corresponds to the constancy of $\delta N(N_c, \varphi(N_*))$ which
was mentioned below Eq.~(\ref{nl1}).

\subsection{Non-linearity generated in scalar dominant phase}
\label{expression}

In order to evaluate Eq.~(\ref{fnl}), we need to know $N_{aF}$ and $N_{abF}$.
If one takes into account the evolution of the curvature perturbations after the scalar dominant phase, 
isocurvature perturbations may remain during preheating/reheating era after inflation.
In such cases, in order to calculate $N_{aF},~N_{abF}$, 
we need to investigate the evolution of the background $e$-folding
number with the effect of short wavelength/radiation component,
which is beyond the scope of the present paper. 

Here, let us evaluate $\zeta(N_F)$ on the uniform energy density hypersurface at $N=N_F$, neglecting the later evolution of the curvature perturbations during the period with $N_c > N > N_F$. 
In this case we can obtain explicit forms of $N_{aF}$ and $N_{abF}$, which appeared in Eqs.
(\ref{defN}) and (\ref{fnl}), written in terms of the background
quantities at $N=N_F$.  
On the super-horizon scales, the uniform energy density hypersurface is equivalent to the 
constant Hubble hypersurface.
Then, $\zeta(N_F)$ is evaluated by the time shift $\delta N$ measured from the $H={\rm constatnt}$ hypersurface.
Therefore at $N=N_F$ we have the equation;
\begin{eqnarray}
H \left( \varphi^a (N_F+\zeta(N_F)) \right)=H \bigl( \varphizero^a (N_F) \bigr)~. \label{exp1}
\end{eqnarray}
The Hubble parameter $H$ is given by Eq.~(\ref{friedmann}).
Solving Eq.~(\ref{exp1}) with respect to $\zeta(N_F)$, 
we obtain
\begin{eqnarray}
\zeta(N_F) \approx N_{aF} \delta \varphi^a_{F} + \frac{1}{2} N_{abF} \delta \varphi^a_{F} \delta \varphi^b_{F}, \label{exp2}
\end{eqnarray}
with
\begin{eqnarray}
&&N_{aF}=-\frac{H_a(\varphi)}{H_c(\varphi) F^c(\varphi)}\biggr|_{\varphi=\varphizero (N_F)}~, \label{exp4} \\
&&N_{abF}=-\frac{U_{ab}(\varphi)}{H_c (\varphi)F^c(\varphi)}\biggr|_{\varphi=\varphizero (N_F)}~, \label{exp5}
\end{eqnarray}
where
\begin{eqnarray}
&&U_{ab}=H_{ab}-\frac{2H_a}{H_d F^d} \left( H_c P^{c}_{~b}+ F^c H_{cb} \right) \nonumber \\
&& \hspace{10mm}-\frac{2H_a H_b}{\left(H_e F^e\right)^2 } \left( F^c H_{cd} F^d
+F^c P^{~d}_{c} H_d \right)~. \nonumber\\ \label{exp3}
\end{eqnarray}
and $H_a \equiv \partial H/ \partial \varphi^a,~H_{ab} \equiv \partial^2 H/ \partial \varphi^a \partial \varphi^b $. 
The explicit forms of $H_{a}$ and $H_{ab}$ are 
shown in appendix \ref{specific}.
The right-hand sides of Eqs.~(\ref{exp4}) and (\ref{exp5}) are written in terms of local quantities,
i.e.,
those depending only on $\phi^I$ and $d\phi^I/dN$ at $N=N_F$.
Hence, once we specify a central background trajectory $\varphizero^a$,
we can readily determine $N_{aF},~N_{abF}$.
$\delta N$ evaluated using the determined $N_{aF}$ and $N_{abF}$ depends 
on $N_F$ unless $N_F > N_c$, where
$N_c$ is a time when the complete convergence of the background trajectories occurs.

\section{Double inflation model with large mass ratio}
\label{example}

Substituting Eqs.~(\ref{exp4}) and (\ref{exp5}) to Eq.~(\ref{fnl}),
the non-linear parameter, $f_{NL}$, can be also defined as a function of the final time, $N_F$.
As mentioned above, the curvature perturbations on uniform energy density hypersurface, $\zeta$, 
evolve when $N_F < N_c$, and hence $f_{NL}$ also evolves.

 In this section, using the formulation given in the previous section \ref{formulation},
 we calculate the non-linear parameter $f_{NL}(N_F)$ for
the double inflation models, which violate the slow-roll conditions
for a certain period when the cosmologically relevant scales are well outside the horizon.
The potential is given by \cite{Silk:1986vc,Turner:1986te},
\begin{eqnarray}
V(\phi,\chi) = {1 \over 2}m_\phi^2\phi^2 + {1 \over 2}m_\chi^2\chi^2~.
\label{doublepotential}
\end{eqnarray}
We assume a large mass ratio, i.e., $m_\chi/m_\phi \gg 1$.
Because of $m_\chi/m_\phi \gg 1$, 
the energy density of $\chi$-field,
$\rho_\chi$, 
decays faster than that of $\phi$-field, $\rho_\phi$.
Hence if $\rho_\phi \gg \rho_\chi$ at initial time, 
 $\rho_\chi$ never dominates the energy density during the later
 evolution of the universe.
In such a model, 
only a single chaotic inflation induced by $\phi$-field occurs. 
It is known that in a single scalar slow-roll inflation,
$f_{NL}$ is suppressed by the slow-roll parameters.
Here we assume the opposite, $\rho_\chi \gg \rho_\phi$, at the initial time. 
In this case the slow-roll conditions are badly violated when 
$\phi - \chi$ equality ($\rho_\phi \simeq \rho_\chi$) occurs.
We assume  that both fields are in the slow-roll phase at the initial time
and the inflation induced by $\phi$-field also occurs after $\phi - \chi$ equality. 
Denoting the initial values of fields at $N=N_*$ as $\phi_*$ and $\chi_*$, 
this assumption implies $\phi_*,\chi_* \gg 1$.

There are several works in which
the primordial non-Gaussianity in the two-scalar chaotic inflation model whose potential is given by Eq.~(\ref{doublepotential}) was investigated.
In Ref.~\cite{Kim:2006te,Vernizzi:2006ve},
the authors analyzed the primordial non-Gaussianity generated during
${\cal N}$-flation~\cite{Dimopoulos:2005ac} model
and gave a simple analytic formula using the slow-roll approximation
even on super-horizon scales based on $\delta N$ formalism.
In Ref.~\cite{Vernizzi:2006ve}, using their slow-roll formula
the authors also calculated the non-linear parameter $f_{NL}$ for
the two-scalar chaotic inflation model whose potential is given by
Eq.~(\ref{doublepotential}) with the mass ratio $m_\phi/m_\chi=9$.
In such case, the slow-roll conditions are not violated during
scalar dominant phase, so they could use the slow-roll formula.
They compared the result obtained by using their analytic formula
with that obtained numerically,
and they found that the non-linear parameter does not become large
in such case.
In Ref.~\cite{Rigopoulos:2005us}, the authors also provided
another formulation for the non-linearity generated on super-horizon scales without slow-roll conditions.
They also calculated the non-linear parameter using their formulation
 for the potential given by Eq.~(\ref{doublepotential}) with
 the mass ratio $m_{\phi}/m_{\chi}=12$.    
They also found that the non-linear parameter evaluated at the end of inflation does not become large
in such model.

Here, using the formulation given in the previous section,
we numerically calculated $f_{NL}$ for the mass parameters 
$m_\phi=0.05,~m_\chi=1.0$ and compared the result with the previous work~\cite{Rigopoulos:2005us}.

\subsection{Numerical calculation for $m_\chi/m_\phi = 20$}
\label{nume}
\begin{figure}[htbp]
  \begin{center}
    \includegraphics{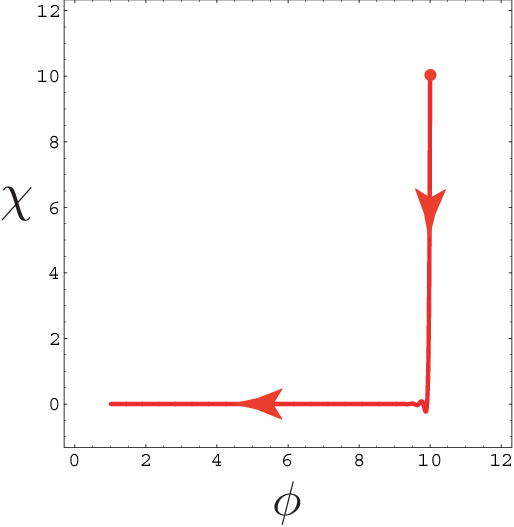}
  \end{center}
  \caption{This figure shows a background trajectory in field space. 
  We set initial values of fields to $\phi_*=\chi_*=10$.}
  \label{trajectory}
\end{figure}

\begin{figure}[htbp]
  \begin{center}
    \includegraphics{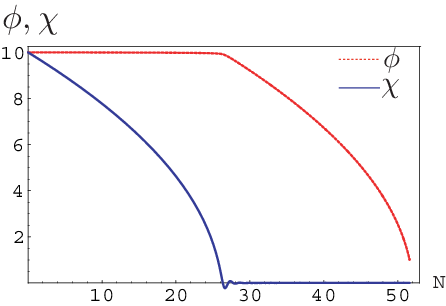}
  \end{center}
  \caption{(color online) This figure shows evolution of $\phi$ (dashed red line) and $\chi$ (solid blue line). 
  We choose the $e$-folding number, $N$, as a time coordinate and set the initial time, $N_*$, to $0$.}
  \label{fig:phi1.eps}
\end{figure}

\begin{figure}[htbp]
  \begin{center}
    \includegraphics{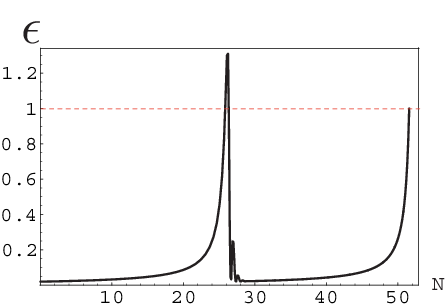}
  \end{center}
  \caption{This figure shows the evolution of the slow-roll parameter $\epsilon$ defined by Eq.~(\ref{epsilon}).}
  \label{fig:epsilon.eps}
\end{figure}

\begin{figure}[htbp]
  \begin{center}
    \includegraphics{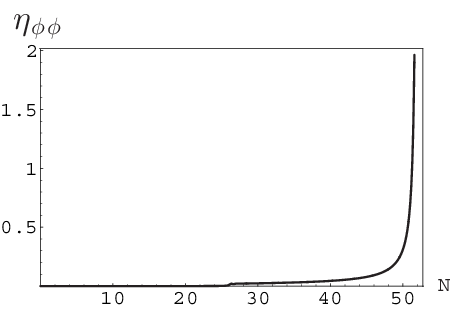}
  \end{center}
  \caption{This figure shows the evolution of the slow-roll parameter $\eta_{\phi\phi}$ defined by Eq.~(\ref{eta}).}
  \label{fig:eta11.eps}
\end{figure}

\begin{figure}[htbp]
  \begin{center}
    \includegraphics{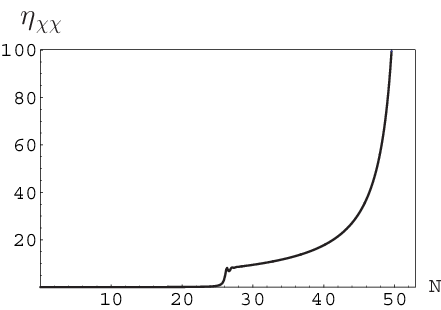}
  \end{center}
  \caption{This figure shows the evolution of the slow-roll parameter $\eta_{\chi\chi}$ defined by Eq.~(\ref{eta}).}
  \label{fig:eta22.eps}
\end{figure}

Using the formulation given in the previous section,
we numerically calculated $f_{NL}$ for the mass parameters 
$m_\phi=0.05,~m_\chi=1.0$.
Hence the mass ratio is $m_\chi/m_\phi=20$.
Initial value of fields are set to $\phi_*=\chi_*=10$.
We choose the $e$-folding number as a time coordinate and set the initial time $N_*$ to $0$.
The central background trajectory in the field space is shown 
in Fig~\ref{trajectory} and the evolution 
of these two fields as a function of $N$ is shown in Fig~\ref{fig:phi1.eps}.
We define slow-roll parameters as
\begin{eqnarray}
&&\epsilon \equiv -{1 \over H}{dH \over dN}~,\label{epsilon}\\
&&\eta_{\phi\phi} \equiv {m_\phi^2 \over V}~,~~ \eta_{\chi\chi} \equiv {m_\chi^2 \over V}~. \label{eta}
\end{eqnarray}
The evolution of the slow-roll parameters is shown in Figs.~\ref{fig:epsilon.eps},~\ref{fig:eta11.eps} and \ref{fig:eta22.eps}.


\begin{figure}[htbp]
  \begin{center}
    \includegraphics{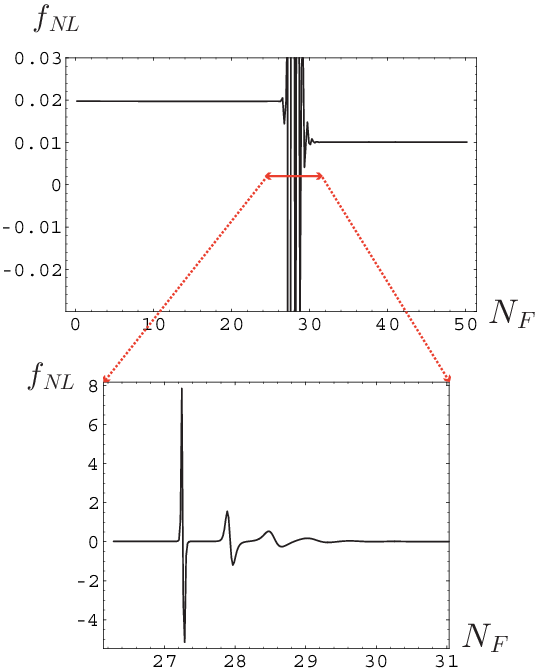}
  \end{center}
  \caption{This figure shows the evolution of non-linear parameter, $f_{NL}(N_F)$. 
After the second inflation, $N_F>30$, the background trajectories in phase space have completely converged. Thus, the curvature perturbation on a constant Hubble hypersurface, $\zeta(N_F)$, remains constant on large scales at $N_F > 30$ and the $f_{NL}(N_F)$ also remains constant in this era.}
  \label{fig:fNL1.eps}
\end{figure}

From Fig.~\ref{fig:phi1.eps}, we see that initially $\chi$-field decays rapidly while
$\phi$-field remains almost at its initial value due to large 
Hubble friction.
At around $N=26$,
$\chi$-field reaches zero and starts damped oscillation.
As shown in Fig.~\ref{fig:epsilon.eps}, around this time,
the slow-roll parameter $\epsilon$ exceeds $1$ for a moment
and  the inflation ends once. From Figs.~\ref{fig:eta11.eps} and \ref{fig:eta22.eps},
also $\eta_{\chi\chi}$ exceeds $1$ after $\chi$-field settles to the minimum,
while,
$\phi$-field is slow-rolling during this phase.

In calculating the non-linear parameter $f_{NL}$,
we regard $f_{NL}$ as a function of $N_F$ 
in the same sense as $\zeta(N_F)$ in Eq.~(\ref{exp2}).
We show the evolution of the non-linear parameter $f_{NL}(N_F)$ in Fig.~\ref{fig:fNL1.eps}.
From this figure, we find that the non-linear parameter $f_{NL}(N_F)$ temporarily oscillates with the maximum amplitude reaching $\sim 8$.
As soon as $\chi$-field settles to its minimum,
the universe becomes dominated by $\phi$-field due to
the decay of $\rho_\chi$, and the universe starts the second 
inflation driven by $\phi$-field.
After the second inflation starts,
iso-curvature perturbation decays rapidly due to the decay of $\chi$.
Hence the time $N_c$ is
before the end of the second inflation in this model.
Then, the sum of $f_{NL}(N_F > 30)$ and the sub-horizon contributions
gives the primordial non-Gaussianity, independent of the reheating process.
The final value of $f_{NL}(N_F>30)$ is $0.01004$.
Hence the generation of large non-Gaussianity due to the
violation of the slow-roll conditions in this model does not occur.

In Ref.~\cite{Rigopoulos:2005us}, the authors numerically calculated the non-linear parameter for the
model whose potential is given by Eq.~(\ref{doublepotential}) with mass ratio
$m_{\phi}/m_{\chi}=12$.
They used their own formalism, which is different from $\delta N$ formalism used in this paper.
The behavior of the non-linear parameter $f_{NL}$ obtained by their calculation
seems similar to our result shown in Fig.~\ref{fig:fNL1.eps}.

\subsection{Approximate analytical expression for $f_{NL}$}
In the previous subsection (\ref{nume}),
by numerical calculation,
we found that the final value of $f_{NL}$ is much less than $1$, although the curvature perturbation becomes
highly non-Gaussian ($f_{NL} \simeq 8$) for a moment.
We can derive the analytical expression for the final value 
of $f_{NL}$ in the limit of large mass ratio $m_\chi/m_\phi \gg 1$,
which is the subject of this subsection.

The background equations are given by 
\begin{eqnarray}
&&H{d \over dN}\left(H{d \phi \over dN}\right) + 3H^2{d\phi \over dN} + m_\phi^2 \phi = 0~, \\
&&H{d \over dN}\left(H{d \chi \over dN}\right) + 3H^2{d\chi \over dN} + m_\chi^2 \chi = 0~, \label{chi}\\
&&H^2 ={
m_\phi^2 \phi^2 + m_\chi^2 \chi^2 \over 
6+\left(\displaystyle {d \phi \over dN}\right)^2 
+ \left(\displaystyle {d \chi \over dN}\right)^2} ~.
\end{eqnarray}
Within the slow-roll approximation, 
these equations reduce to
\begin{eqnarray}
&&{d \over dN}{\phi} \simeq -{m_\phi^2 \phi \over 3H^2}~,~~~~{d \over dN}{\chi} \simeq - {m_\chi^2 \chi \over 3H^2}~, \nonumber \\
&&H^2 \simeq {1 \over 6}\left(m_\phi^2 \phi^2 + m_\chi^2 \chi^2\right)~.
\label{sbe}
\end{eqnarray}
At the initial time, $N=N_*$, 
both fields are in the slow-roll phase and $\phi_{*} \simeq \chi_{*} \gg 1$. 
Then the ratio of the time derivatives of the scalar fields is 
\begin{eqnarray}
{d\phi/dN \over d\chi/dN} \simeq \frac{m_\phi^2}{m_\chi^2} {\phi \over \chi} \ll 1~,
\end{eqnarray}
and also $\rho_\chi \gg \rho_\phi$ in the limit of large mass ratio. 
Hence,
to a good approximation, 
$\phi(N) \simeq \phi_{*}$ until the energy density of $\phi$-field dominates the total energy density of the universe,
i.e., $\rho_\phi \simeq \rho_\chi$. 
On the other hand, 
solving Eq.~(\ref{sbe}), the evolution of $\chi$-field during the slow-roll phase
can be obtained as
\begin{eqnarray}
\chi(N) \simeq \sqrt{\chi_{*}^2 - 4N}~. 
\label{slc}
\end{eqnarray}
Here we recall that we have set $N_* = 0$. 
After the time when $\chi \sim 1$, 
$\chi$-field is no longer in the slow-roll phase and 
$\rho_\chi$ evolves as;
\begin{eqnarray}
\rho_\chi(N) &\simeq& \rho_\chi(\chi(N_{os})) \exp \left[-3(N - N_{os})\right]~, 
\label{osc}
\end{eqnarray}
where $N_{os}$ represents the time when $\chi$-field starts oscillation.
From Eq.~(\ref{slc}), 
we have
\begin{eqnarray}
N_{os} \simeq {\chi_{*}^2 \over 4} - {{\chi(N_{os})}^2 \over 4}~.
\end{eqnarray} 
Then, Eq.~(\ref{osc}) can be rewritten as
\begin{eqnarray}
\rho_\chi(N) \simeq \tilde{\rho}_\chi \exp \left( -3 N + {3 \over 4}\chi_{*}^2\right)~, 
\label{osc2}
\end{eqnarray}
where $\tilde{\rho}_\chi$ is independent of $N$, $\chi_{*}$ and $\phi_*$.
Since $\rho_\chi$ decays in time for $N>N_{os}$, $\rho_\phi$ dominates the total energy density of the universe at some time
and $\phi$-field starts slow-rolling.
Let us denote the $e$-folding number at the time when $ \rho_\phi \simeq \rho_\chi$ as $N_{eq}$. 
For $N > N_{eq}$, the evolution of $\phi$-field is given by
\begin{eqnarray}
\phi(N) \simeq \sqrt{\phi_{*}^2 - 4(N-N_{eq})}~,
\label{slp}
\end{eqnarray}
during the slow-roll phase. 
$N_{eq}$ is obtained from the equation,
\begin{eqnarray}
\rho_\phi(N_{eq}) \simeq {1 \over 2}m_{\phi}^2 \phi(N_{eq})^2 \simeq \rho_\chi(N_{eq})~.
\label{equality}
\end{eqnarray}
Combining  Eq.~(\ref{osc2}) with Eq.~(\ref{equality}), we have 
\begin{eqnarray}
N_{eq} \simeq {\chi_{*}^2 \over 4} - {1 \over 3}\log{m_\phi^2 \phi_{*}^2 \over 2 \tilde{\rho}_\chi}~.
\label{neq}
\end{eqnarray}
The total energy density after $N = N_{eq}$ is approximately $\rho_\phi$. 
Hence, 
using Eqs.~(\ref{slp}) and (\ref{neq}), 
we have
\begin{eqnarray}
\rho(N) &\simeq &
\rho_\phi(N) \nonumber\\
 &\simeq& {m_\phi^2 \over 2}\phi^2(N) \nonumber\\
&\simeq& {m_\phi^2 \over 2}\left[\phi_{*}^2+\chi_{*}^2- 4 \left(N 
+ {1 \over 3} \log{m_\phi^2 \phi_{*}^2 \over 2 \tilde{\rho}_\chi}\right)\right]~.\nonumber\\
\label{rhon}
\end{eqnarray}

After $N=N_{eq}$, 
the universe is dominated by the single slow-roll component
$\phi$-field, and the curvature perturbation on the uniform density 
hypersurface becomes constant in time on super-horizon scales. 
From Eq.~(\ref{rhon}), 
we obtain an expression for the $e$-folding number 
at the final time in terms of $\phi_{*}$ and $\chi_{*}$ as
\begin{eqnarray}
N(N_F,\phi_{*},\chi_{*})& \simeq& \frac{1}{4} \left( \phi_{*}^2 + \chi_{*}^2 \right)
- {2 \over 3}\log \phi_{*} \nonumber\\
&&
 - {\rho_F \over 2m_\phi^2}
- {1 \over 3}\log {m_\phi^2 \over 2 \tilde{\rho}_\chi}~, \label{efolddouble}
\end{eqnarray}
where $\rho_F = \rho(N_F)$.
This gives
\begin{eqnarray}
-{6 \over 5}f_{NL}(N_F) &\simeq& { N_{I}N_{J}N^{IJ} \over \left(N_KN^K\right)^2} \nonumber\\
&=&{2 \over \phi_{*}^2 + \chi_{*}^2}
\left(1 + {\cal O}(\phi_{*}^{-2},~\chi_{*}^{-2} ) \right), \label{difnl}
\end{eqnarray}
where $N_I = \partial N / \partial \phi^I_{*}$, 
$N_{IJ} = \partial^2 N / \partial \phi^I_{*}\partial \phi^J_{*}$ $($ Here, $\phi^I = (\phi, \chi))$.
Hence $f_{NL}$ is ${\cal O}(\phi_{*}^{-2},~\chi_{*}^{-2} )$,
and it is much less than $1$.
$f_{NL}$ estimated using this expression is $\approx 0.01$ for $m_\phi=0.05,~m_\chi=1.0$
and $\phi^{*}=\chi^{*}=10$,
which agrees well with the numerical result in the previous
subsection (see \ref{nume}).

The reason why we get small $f_{NL}$ in the double 
inflation model can be understood as follows.
In the case of a large mass ratio,
evolution of the universe can be clearly divided
into three stages.
First one is the inflation induced almost by a single field,
i.e., $\chi$-field.
The second one is the non-inflationary phase where
the energy density of the oscillating $\chi$-field
is still larger than that of the $\phi$-field.
The third one is the inflation induced by a single field,
i.e., $\phi$-field.
Then the total $e$-folding number during all these stages 
mainly comes from the two inflationary stages,
i.e.,
the first and the third stages.
The perturbation of the $e$-folding number generated during the second stage is
negligible.
Hence as a crude approximation,
the dynamics can be regarded as just the sum of 
two single field inflationary phases.
This picture gives the first term of Eq.~(\ref{efolddouble}).
The correction to this picture is represented by the second term,
which is indeed minor for $N \simeq 60$.

The derivation of Eq.~(\ref{difnl}) can be straightforwardly
extended to more general double inflation models with 
\begin{eqnarray}
V(\phi,\chi) = c_1 \phi^{2p} + c_2 \chi^{2p},
\end{eqnarray}
where $p$ is an arbitrary positive integer.
In this case,
the final value of $f_{NL}$ is more suppressed than 
Eq.~(\ref{difnl}) by a factor $1/p$.

\subsection{Comment on $f_{NL}$ in ${\cal N}$-flation model}
Discussion of the previous subsection about the small
$f_{NL}$ in the double inflation model gives us some
insight into the generation of the non-Gaussianity in the 
so-called ${\cal N}$-flation model~\cite{Dimopoulos:2005ac}, in which the potential is given by
\begin{eqnarray}
V(\phi)=\sum_{I=1}^{\cal N} \frac{m_I^2}{2} {\phi^I}^2.
\end{eqnarray}
This is a generalization of the double inflation model
to an arbitrary number of fields.
We consider the case in which all fields have large initial amplitude, i.e.,
$\phi^I \gg 1$. We also assume $m_\phi < m_\chi < \cdots < m^{\cal N}$.
In this case,
$\phi^{\cal N}$ field decays first.
If $\rho_{\cal N}$ dominates the total energy density of the
universe until $\phi^{\cal N} \sim 1$ 
\footnote{If $\rho_{\cal N}$ becomes subdominant before
$\phi^{\cal N} \sim 1$,
then $\phi^{\cal N}$ 
is slowly rolling until $H$ decreases to $m_{\cal N}$.
Such a case was studied in \cite{Kim:2006te,Vernizzi:2006ve}
under the slow-roll and horizon crossing approximation.
The final value of $f_{NL}$ was found to be suppressed 
by the slow-roll parameters.},
then the inflation is almost like the single field inflation
by $\phi^{\cal N}$-field.
After $\phi^{\cal N}$-field decays,
then $\phi^{{\cal N}-1}$-field decays and so on.
Hence the leading $e$-folding number would be given by
\begin{eqnarray}
N \simeq \frac{1}{4} \sum_{I=1}^{\cal N} {\phi^I_{*}}^2.
\end{eqnarray}
Then the corresponding $f_{NL}$ is
\begin{eqnarray}
-{6 \over 5}f_{NL} \sim 2 {\left( \sum_{I=1}^{\cal N} {\phi^I_{*}}^2 \right)}^{-1}
\sim {1 \over 2N}.
\end{eqnarray}
Roughly speaking,
$f_{NL}$ is suppressed by the inverse of the $e$-folding number.
This result is quite similar to that in Ref.~\cite{Kim:2006te}, where the authors studied under the slow-roll approximation.

\section{Summary}
\label{summary}
Based on the $\delta N$ formalism,
we have derived a useful formula for calculating 
the primordial non-Gaussianity due to the super-horizon evolution of the curvature perturbation in multi-scalar inflation
without imposing slow-roll conditions.
This formula can apply for the inflation models with general field space metric, $h_{IJ}$, as far as super-horizon contributions are concerned. 
Generally, 
when one calculates the non-Gaussianity of the curvature perturbations, 
one has to solve the second order perturbation equations. In doing so 
for a multi-scalar inflation, there appear 
tensorial quantities with respect to the indices of the field components. 
Our formula reduces the problem of calculating the non-linear parameter $f_{NL}$
to solving only first order perturbation equations 
for two vector quantities.
This reduces ${\cal O}({\cal N}^2)$ calculations to ${\cal O}({\cal N})$
ones where ${\cal N}$ is the number of the scalar field components.
Hence our formalism has a great advantage for the numerical
evaluation of $f_{NL}$ in the inflation model composed of a large
number of fields.
 
We have also studied the primordial non-Gaussianity in double inflation 
model as an example that violates slow-roll conditions by using our formalism.
We found that, although $f_{NL}$ defined for the curvature perturbation
on a constant Hubble hypersurface exceeds $1$ for a moment
around the time when the slow-roll conditions are violated,
the final value of $f_{NL}$ is suppressed by the slow-roll 
parameters evaluated at the time of horizon exit.
We have shown that 
this can be understood even analytically in the $\delta N$ formalism.
This result is straightforwardly extended to more general double 
inflation model and ${\cal N}$-flation model.

\begin{acknowledgments}
SY thanks Takashi Nakamura for stimulating comments. We would like to thank
F. Vernizzi and G. I. Rigopoulos for useful comments
and P. Shukla for pointing out typos.
TT is supported 
by Monbukagakusho Grant-in-Aid
for Scientific Research Nos.~17340075 and~19540285. 
This work is also supported in part by the 21st Century COE 
``Center for Diversity and Universality in Physics'' at Kyoto
 university, from the Ministry of Education,
Culture, Sports, Science and Technology of Japan. 
The authors thank the Yukawa Institute for Theoretical Physics at Kyoto University. Discussions during YITP workshop YITP-W-07-10 on "KIAS-YITP Joint Workshop, String Phenomenology and Cosmology"
were useful to complete this work.

\end{acknowledgments}

\appendix
\section{specific expression}
\label{specific}
In this appendix, we show the explicit forms of $P^a_{~b}(=P^{Ij}_{iJ})$ in Eq.~(\ref{Pdefined}),
$Q^a_{~bc}(=Q^{Ijk}_{iJK})$ in Eq.~(\ref{Qdefined}),  
$H_a(=H^i_I)$ in Eq.~(\ref{exp2}) and $H_{ab}(=H^{ij}_{IJ})$ in Eq.~(\ref{exp3}).
$P^{Ij}_{iJ}$ is given by
\begin{eqnarray}
&&P^{I1}_{1J}=0~,\quad P^{I2}_{1J}=\delta^I_{~J}~,\nonumber\\
&&P^{I1}_{2J}=-{V \over H^2}\left({V^I_{~J} \over V} - {V^IV_J \over V^2}\right)~,\nonumber\\ 
&&P^{I2}_{2J}=\varphi^I_2\varphi_{2J} + {V^I \over V}\varphi_{2J} -{V
 \over H^2}\delta^I_{~J}~. 
\label{PIJ}
\end{eqnarray}
$Q^{Ijk}_{iJK}$  by
\begin{eqnarray}
&&\hspace*{-5mm}
Q^{I11}_{1JK} = Q^{I21}_{1JK}=Q^{I12}_{1JK}=Q^{I22}_{1JK}=0~,\nonumber\\
&&\hspace*{-5mm}
Q^{I11}_{2JK} = -{V \over H^2}\Biggl({V^I_{~JK} \over V} -{V^I_{~J}V_K \over V^2} \nonumber\\
&&\qquad\qquad -{V^I_{~K}V_J \over V^2}
 -{V^IV_{JK} \over V^2} + 2{V^IV_{J}V_K \over V^3} \Biggr)~,\nonumber\\
&&\hspace*{-5mm}
Q^{I12}_{2JK} = \left({V^I_{~J} \over V} - {V^IV_J \over V^2}\right)\varphi_{2K}~,\nonumber\\
&&\hspace*{-5mm}
Q^{I21}_{2JK} = \left({V^I_{~K} \over V} - {V^IV_K \over V^2}\right)\varphi_{2J}= Q^{I12}_{2KJ}~,\nonumber\\
&&\hspace*{-5mm}
Q^{I22}_{2JK} = \delta^I_{~K}\varphi_{2J} + \delta^I_{~J}\varphi_{2K} + \delta_{JK} \left( \varphi^I_2+{V^I \over V}\right)~.
\label{QIJK}
\end{eqnarray}
$H_I^i$ is given by
\begin{eqnarray}
H_I^1={H \over 2}{V_I \over V}~,~~H_I^2={H^3 \over 2V}\varphi_{2I}~,\label{H}
\end{eqnarray}
and $H_{IJ}^{ij}$ by
\begin{eqnarray}
&&H_{IJ}^{11}={H \over 2}\left({V_{IJ} \over V} - {V_IV_J \over 2V^2}\right)~, \nonumber\\
&&H_{IJ}^{12}={H^3 \over 4V^2}V_I\varphi_{2J}~,~~H_{IJ}^{21}={H^3 \over 4V^2}V_J\varphi_{2I}~,
\nonumber\\
&&H_{IJ}^{22}={H^3 \over 2V}\left(\delta_{IJ}+{3H^2 \over 2V}\varphi_{2I}\varphi_{2J}\right)~.
\end{eqnarray}

\section{Extension to the general field space metric}
\label{hij}

In this appendix,
we will extend the formulation given in sec.~II for the flat field space metric $h_{IJ}=\delta_{IJ}$ to the general case. 
For general case,
the background equations corresponding to Eq.~(\ref{back1}) 
for flat field space are given by
\begin{eqnarray}
\frac{d\varphi^I_1}{dN}=F^I_1~,~~\frac{D \varphi^I_2}{dN}=F^I_2~,
\label{curveback}
\end{eqnarray}
where $DA^I \equiv dA^I + \Gamma^I_{~JK}A^J d\varphi^K_1$, 
$\Gamma^I_{~JK}$ is Christoffel symbol
with respect to the field space metric $h_{IJ}$, and
\begin{eqnarray}
F^I_1 = \varphi^I_2~,~~~~F^I_2 = -\frac{V}{H^2} \left( \varphi^I_2+h^{IJ} \frac{V_J}{V} \right)~.
\label{curveF}
\end{eqnarray}
In the same way as Eq.~(\ref{per1}), we can define the perturbation as 
\begin{eqnarray}
\delta \varphi^a (N) =\varphi^a( \lambda+ \delta \lambda;
 N)-\varphi^a(\lambda; N)~, 
\label{ape1}
\end{eqnarray}
and we decompose $\delta \varphi^I_i$ as
\begin{eqnarray}
\delta \varphi^a = \delta \varphione^a + {1 \over 2} \delta \varphitwo^a~.
\end{eqnarray}
The curvature perturbation can be expanded in terms of $\delta \varphi^I_i$ as
\begin{eqnarray}
\zeta &\simeq& N_a\delta \varphi^a + {1 \over
 2}N_{ab}\delta\varphi^a\delta\varphi^b + \cdots~.
\label{appdeltaN}
\end{eqnarray}
However,
$\delta \varphi^I_i$ defined in this way does not
transform as a vector under the "coordinate transformation",
$\phi^I \to {\bar \phi}^I(\phi)$ because Eq.~(\ref{ape1})
takes the difference of variables at different points in field space.
Hence the evolution equation for $\delta \varphi^I_i$ lacks
the manifest covariance.
To avoid this drawback (though not essential),
we introduce new perturbation variables
$\delta\tilde\varphi^a=\delta\tvarphione{}^a+{1\over
2}\delta\tvarphitwo^a$ 
defined by 
\begin{eqnarray}
&&\delta\tvarphione^I_{1} \equiv {d \varphi^I_1 \over d\lambda}\delta \lambda~,~~
\delta\tvarphitwo^I_{1} \equiv {D \over d\lambda}{d \varphi^I_1 \over d\lambda}\delta \lambda^2~, \\
&&\delta\tvarphione^I_{2} \equiv {D \varphi^I_2 \over d\lambda}\delta \lambda~,~~
\delta\tvarphitwo^I_{2} \equiv {D^2 \varphi^I_2 \over d\lambda^2}\delta \lambda^2~, 
\end{eqnarray}
which transform as
vectors.
Differentiating the background equations~(\ref{curveback}) with
respect to $\lambda$, we obtain the evolution equations for the
perturbations of the scalar fields in curved field space metric.
To the first order, we have
\begin{eqnarray}
{D \over dN}\delta\tvarphione^a(N) = \left(P^{a}_{~b}(N) + \Delta
					  P^{a}_{~b}(N)\right)
\delta\tvarphione^b(N)~,
\end{eqnarray}
where the $P^{a}_{~b}$ is given by
\begin{eqnarray}
P^{a}_{~b} = {D F^a \over \partial \varphi^b}\biggr|_{\varphi=\varphizero (N)}
~,
\label{curveP}
\end{eqnarray}
and $\Delta P^{a}_{~b}$, which vanishes in the flat 
field space metric representing terms due to the curvature of the field
space, is given by
\begin{eqnarray}
&&\Delta P^{I1}_{1J}=\Delta P^{I2}_{1J}=\Delta P^{I2}_{2J}=0~, \nonumber\\
&&\Delta P^{I1}_{2J}=-R^{I}_{~LJK}\varphi^L_{2}\varphi^K_{2}~.
\end{eqnarray}
Here, $R^I_{~LJK}$ represents Riemann tensor associated with the field space metric $h_{IJ}$.
Evolution equation for $\delta\tvarphitwo^a$ is also given by
\begin{eqnarray}
{D \over dN}\delta\tvarphitwo^a(N) 
&\!\! =&\!\! \left(P^a_{~b}(N) + \Delta P^{a}_{~b}(N)\right)\delta\tvarphitwo^b(N) \nonumber\\
&& +\left(Q^{a}_{~bc}(N) + \Delta Q^{a}_{~bc}(N)\right)\delta \tvarphione^b(N)\delta
\tvarphione^c(N)~,\nonumber\\
\end{eqnarray}
where
\begin{eqnarray}
&&Q^{a}_{~bc}= {D^2 F^a \over \partial \varphi^b \partial \varphi^c}\biggr|_{\varphi=\varphizero (N)} 
= {D P^a_{~b} \over \partial \varphi^c}\biggr|_{\varphi=\varphizero (N)}~,
\end{eqnarray}
and the explicit form of $\Delta Q^{a}_{~bc}$ is given by
\begin{eqnarray}
&&\Delta Q^{I11}_{1JK} = -R^{I}_{~JKL}\varphi^L_{2}~,\nonumber\\
&&\Delta Q^{I12}_{1JK} = \Delta Q^{I21}_{1JK} = \Delta Q^{I22}_{1JK} = 0~, \nonumber\\
&&\Delta Q^{I11}_{2JK} = -\nabla_KR^{I}_{~MJL}\varphi^M_2\varphi^L_2~, \nonumber\\
&&\Delta Q^{I12}_{2JK} = -2R^{I}_{~KJL}\varphi^L_2~,~~
\Delta Q^{I21}_{2JK} = -R^{I}_{~LKJ}\varphi^L_2~, \nonumber\\
&&\Delta Q^{I22}_{2JK} = 0~,
\end{eqnarray}
where $\nabla_K$ is covariant derivative with respect to $h_{IJ}$.

By introducing the matrix $\Lambda$ as the solution of
\begin{eqnarray}
{D \over dN}\Lambda^a_{~b}(N,N_*) = \left(P^{a}_{~c}(N)+\Delta P^{a}_{~c}(N)\right)\Lambda^{c}_{~b}(N,N_*),\nonumber\\
\end{eqnarray}
with the condition $\Lambda^{Ij}_{iJ}(N,N)=\delta^I_{\, J} \delta^{~ i}_j$,
formal solutions of the first and second order equations
are given by
\begin{eqnarray}
&&\delta\tvarphione^a(N) = \Lambda^a_{~b}(N,N_*)
\delta \tvarphione^b(N_*), \nonumber\\
&&\delta \tvarphitwo^a(N) =\int^N_{N_*}dN'
 \Lambda^a_{~b} (N,N')
\nonumber\\
&&\qquad \times\left[Q^{b}_{~cd}(N')
+ \Delta Q^{b}_{~cd}(N')\right]\delta\tvarphione^c(N')
\delta\tvarphione^d(N')~.\nonumber\\ 
\end{eqnarray}

The expression for the curvature perturbation in terms of $\delta
\varphi^a$, Eq.~(\ref{appdeltaN}), can be also 
rewritten as that in terms of $\delta \tilde\varphi{}^a$. 
To make the expression more concise, 
we introduce new variables $\tilde{N}_{a}$ and $\tilde{N}_{ab}$ as
expansion coefficients of $\zeta$ in terms of $\delta\tilde \varphi{}^a$
so that we have 
\begin{eqnarray}
\zeta &\simeq& \tilde{N}_{aF}\delta \tilde{\varphi}{}^a_{F} 
+ {1 \over 2} \tilde{N}_{abF} \delta \tilde{\varphi}{}^a_{F} \delta
\tilde{\varphi}{}^b_{F}
+\cdots~. 
\label{appdeltaN2}
\end{eqnarray}
Comparing the two expressions (\ref{appdeltaN}) and (\ref{appdeltaN2}), 
we find 
\begin{eqnarray}
&&\tilde{N}_{I}^1 = N_{I}^1 - N_{K}^2\Gamma^K_{~JI}\varphi^J_2~,~~\tilde{N}_{I}^2 = N_{I}^2~, \cr
&&\tilde{N}_{IJ}^{11} = N_{IJ}^{11} + N_{KI}^{22}\Gamma^K_{~ML}\Gamma^{L}_{~NJ}\varphi^M_2\varphi^N_2 \cr
&&\quad\quad+ \left(N_{IK}^{12} + N_{IK}^{21}\right)\Gamma^K_{~LJ}\varphi^L_2 \cr
&&\quad\quad-N_K^2\nabla_{I}\Gamma^K_{LJ}\varphi^L_2- N_{K}^{1}\Gamma^K_{~IJ} + N_{K}^2\Gamma^K_{~IL}\Gamma^{L}_{~NJ}\varphi^N_2~,\cr
&&\tilde{N}_{IJ}^{12} = N_{IJ}^{12} - N_K^{2}\Gamma^K_{~IJ} 
- N_{KJ}^{22}\Gamma^K_{~IL}\varphi^L_2~, \cr
&&\tilde{N}_{IJ}^{21} = N_{IJ}^{21} - N_K^{2}\Gamma^K_{~IJ} 
- N_{KI}^{22}\Gamma^K_{~JL}\varphi^L_2~, \cr
&&\tilde{N}_{IJ}^{22} = N_{IJ}^{22}~.
\end{eqnarray}
From the constancy of $\tilde N_a\delta\tvarphione^a$, 
we find 
\begin{eqnarray}
\tilde {N}_a (N) = \tilde{N}_{bF} \Lambda^b_{~a}(N_F,N)~. 
\end{eqnarray}
We also redefine $\Theta^I_i$
in Eq.~(\ref{deftheta}) as 
\begin{eqnarray}
\Theta^a (N) = \Lambda^{a}_{~b}(N,N_*)A^{bc}\tilde N_{c*}~,
\end{eqnarray}
where $A^{ab}$ is defined by
\begin{eqnarray}
\langle\delta \tilde{\varphi}^a_{*}\delta \tilde{\varphi}^b_{*}\rangle
 = A^{ab}\left({H_* \over 2\pi}\right)^2~.
\end{eqnarray}
$A^{ab}$ for general case also depends on the field space metric, $h_{IJ}$.
 
Then, the expression for the non-linear parameter in general cases is
also given by 
Eq.~(\ref{fnl}) with 
$N_c(N')$ and $Q^c_{~ab}(N')$ replaced by 
$\tilde N_c(N')$ and $Q^c_{~ab}(N')+\Delta Q^c_{~ab}(N')$.


\begin{thebibliography}{99}
\bibitem{Komatsu:2001rj}
  E.~Komatsu and D.~N.~Spergel,
  Phys.\ Rev.\ D {\bf 63}, 063002 (2001)
  [arXiv:astro-ph/0005036].
  
\bibitem{Bartolo:2004if}
  N.~Bartolo, E.~Komatsu, S.~Matarrese and A.~Riotto,
  Phys.\ Rept.\  {\bf 402}, 103 (2004)
  [arXiv:astro-ph/0406398].

\bibitem{:2006uk}
    [Planck Collaboration],
  arXiv:astro-ph/0604069.
 
\bibitem{Babich:2004yc}
  D.~Babich and M.~Zaldarriaga,
  Phys.\ Rev.\  D {\bf 70}, 083005 (2004)
  [arXiv:astro-ph/0408455].
  
\bibitem{Okamoto:2002ik}
  T.~Okamoto and W.~Hu,
  Phys.\ Rev.\  D {\bf 66}, 063008 (2002)
  [arXiv:astro-ph/0206155].
  
\bibitem{Bartolo:2005fp}
  N.~Bartolo, S.~Matarrese and A.~Riotto,
  JCAP {\bf 0508}, 010 (2005)
  [arXiv:astro-ph/0506410].

\bibitem{Kogo:2006kh}
  N.~Kogo and E.~Komatsu,
  Phys.\ Rev.\  D {\bf 73}, 083007 (2006)
  [arXiv:astro-ph/0602099].

\bibitem{Maldacena:2002vr}
  J.~M.~Maldacena,
  JHEP {\bf 0305}, 013 (2003)
  [arXiv:astro-ph/0210603].

\bibitem{Kim:2006te}
  S.~A.~Kim and A.~R.~Liddle,
  Phys.\ Rev.\ D {\bf 74}, 063522 (2006)
  [arXiv:astro-ph/0608186].
  
\cite{Vernizzi:2006ve}
\bibitem{Vernizzi:2006ve}
  F.~Vernizzi and D.~Wands,
  JCAP {\bf 0605}, 019 (2006)
  [arXiv:astro-ph/0603799].

\bibitem{Battefeld:2006sz}
  T.~Battefeld and R.~Easther,
  JCAP {\bf 0703}, 020 (2007)
  [arXiv:astro-ph/0610296].
\bibitem{Battefeld:2007en}
  D.~Battefeld and T.~Battefeld,
  arXiv:hep-th/0703012.
  
\bibitem{Choi:2007su}
  K.~Y.~Choi, L.~M.~H.~Hall and C.~van de Bruck,
  JCAP {\bf 0702}, 029 (2007)
  [arXiv:astro-ph/0701247].
  
\bibitem{Rigopoulos:2005xx}
  G.~I.~Rigopoulos, E.~P.~S.~Shellard and B.~J.~W.~van Tent,
  Phys.\ Rev.\  D {\bf 73}, 083521 (2006)
  [arXiv:astro-ph/0504508].
\bibitem{Rigopoulos:2005ae}
  G.~I.~Rigopoulos, E.~P.~S.~Shellard and B.~J.~W.~van Tent,
  Phys.\ Rev.\  D {\bf 73}, 083522 (2006)
  [arXiv:astro-ph/0506704].
\bibitem{Rigopoulos:2005us}
  G.~I.~Rigopoulos, E.~P.~S.~Shellard and B.~J.~W.~van Tent,
  Phys.\ Rev.\  D {\bf 76}, 083512 (2007)
  [arXiv:astro-ph/0511041].
\bibitem{Seery:2005wm}
  D.~Seery and J.~E.~Lidsey,
  JCAP {\bf 0506}, 003 (2005)
  [arXiv:astro-ph/0503692]. 
\bibitem{Seery:2005gb}
  D.~Seery and J.~E.~Lidsey,
  JCAP {\bf 0509}, 011 (2005)
  [arXiv:astro-ph/0506056].

\bibitem{Yokoyama:2007uu}
  S.~Yokoyama, T.~Suyama and T.~Tanaka,
  arXiv:0705.3178 [astro-ph].

\bibitem{Lyth:2004gb}
  D.~H.~Lyth, K.~A.~Malik and M.~Sasaki,
  JCAP {\bf 0505}, 004 (2005)
  [arXiv:astro-ph/0411220].

\bibitem{Starobinsky:1986fx}
  A.~A.~Starobinsky,
  JETP Lett.\  {\bf 42}, 152 (1985)
  [Pisma Zh.\ Eksp.\ Teor.\ Fiz.\  {\bf 42}, 124 (1985)].
\bibitem{Sasaki:1995aw}
  M.~Sasaki and E.~D.~Stewart,
  Prog.\ Theor.\ Phys.\  {\bf 95}, 71 (1996)
  [arXiv:astro-ph/9507001]. 
\bibitem{Sasaki:1998ug}
  M.~Sasaki and T.~Tanaka,
  Prog.\ Theor.\ Phys.\  {\bf 99}, 763 (1998)
  [arXiv:gr-qc/9801017].

\bibitem{Lyth:2005fi}
  D.~H.~Lyth and Y.~Rodriguez,
  Phys.\ Rev.\ Lett.\  {\bf 95}, 121302 (2005)
  [arXiv:astro-ph/0504045].

\bibitem{Silk:1986vc}
  J.~Silk and M.~S.~Turner,
  Phys.\ Rev.\  D {\bf 35}, 419 (1987).
  

\bibitem{Turner:1986te}
  M.~S.~Turner, J.~V.~Villumsen, N.~Vittorio, J.~Silk and R.~Juszkiewicz,
  Astrophys.\ J.\  {\bf 323}, 423 (1987).
\bibitem{Chen:2006nt}
  X.~Chen, M.~x.~Huang, S.~Kachru and G.~Shiu,
  JCAP {\bf 0701}, 002 (2007)
  [arXiv:hep-th/0605045].

\bibitem{Byrnes:2006vq}
  C.~T.~Byrnes, M.~Sasaki and D.~Wands,
  Phys.\ Rev.\  D {\bf 74}, 123519 (2006)
  [arXiv:astro-ph/0611075].


\bibitem{Dimopoulos:2005ac}
  S.~Dimopoulos, S.~Kachru, J.~McGreevy and J.~G.~Wacker,
  arXiv:hep-th/0507205.

\end{thebibliography}
\end{document}